# Understanding the Effect of Long-Term Memory Model Parameters in Pole-Zero Identification for Stability Analysis of Power Amplifiers

Libe Mori, Aitziber Anakabe, Juan-Mari Collantes and Vincent Armengaud

*Abstract*—Understanding the nature of potential instabilities is indispensable for the stabilization of power amplifiers. Pole-zero identification is one of the techniques that can be used to determine the stability of a design in large-signal operation. In this work, the possible presence of poles at the fundamental frequency linked to the long-term memory parameters of the transistor's model (self-heating and traps) is presented and discussed. The paper shows how their effect on the identified frequency responses around the fundamental frequency may compromise the stability analysis results and the assessment of stability margins. The low observability of the poles at the fundamental frequency highlights the importance of an accurate identification of real poles in low-frequency bands. A specific algorithm for the automatic frequency domain identification of non-resonant frequency responses and a procedure for detecting and reducing overfitting of real poles is proposed in this article. The benefits of the proposed methodology to correctly detect and analyze real poles at low frequencies is demonstrated through Monte-Carlo sensitivity analyses of two different amplifier designs.

## I. INTRODUCTION

ELECTRICAL performances of power transistors are affected by long-term memory effects as self-heating and traps. Reliable large-signal simulations of power amplifiers (PAs) require accurate non-linear compact models that include the modeling of these long-term memory effects. Electro-thermal non-linear models can also be used to detect thermal instabilities as in [1], which can take place in multi-finger bipolar devices. Typical non-linear electro-thermal models compute the junction temperature from the instantaneous dissipated power through a thermal network [2]-[7]. This thermal network is made of one or several RC networks that model the heat exchange from the junction to the case or board. Drain-lag and/or gate-lag trapping, traps' capture and emission time constants are also modeled through RC networks [4]-[11].

Along with an accurate calculation of amplifier performances, large-signal stability analyses are required to detect and prevent undesired oscillations and to determine the regions of stable functioning versus relevant circuit parameters.

A conventional approach for large-signal stability analysis is pole-zero identification of frequency responses obtained by linearizing the circuit around its periodic steady state [12], [13]. Pole-zero identification can also be used as a complementary tool in other stability techniques for complex circuits [14], [15]. Automatic tools for stability analysis through pole-zero identification are available in microwave commercial CAD software [16], [17]. Pole-zero identification has been successfully used to analyze the different types of local bifurcations of periodic solutions in PAs: Flip bifurcations [19], [20], secondary Hopf bifurcations [21], [22] and D-type bifurcations [23].

The condition for D-type bifurcation corresponds to the crossing of a family of complex conjugate poles through the imaginary axis of the complex plane at $\pm jn2\pi f_{in}$, with $n = 0,1,2…$ [18]. In practice, D-type bifurcations analyzed with pole-zero identification are commonly detected through the crossing of a pair of complex conjugate poles crossing at $\pm j2\pi f_{in}$ ($n = 1$), instead of the crossing of a real pole ($n = 0$) [23], [24]. This is because common identification algorithms are more accurate in the detection of resonant peaks in the frequency response by means of pairs of complex conjugate poles than fitting the non-resonant magnitude and phase changes associated with real poles [25].

However, the RC networks modeling long-term memory effects introduce negative real poles in the transfer function with very small values (-$\varepsilon$), in the order of kHz to MHz. The time constants of these real poles are very long compared to the fundamental frequency of the PAs. As the input power at $f_{in}$ increases, Floquet repetitions of these real poles appear at $-\varepsilon \pm j2\pi f_{in}$. As $\varepsilon$ is several orders of magnitude smaller than $f_{in}$,

Manuscript submitted July 19, 2023; revised September 29, 2023; accepted October 18, 2023. This work has been funded by MCIN/AEI/10.13039/501100011033, Grant PID2019-104820RB-I00 and partially funded by the Departamento de Educación del Gobierno Vasco (IT1533-22). *Corresponding author: L. Mori*.

L. Mori is with Departamento de Arquitectura y Tecnología de Computadores, University of the Basque Country UPV/EHU, 20018 Donostia, Spain (e-mail: libe.mori@ehu.eus).

A. Anakabe and J.M. Collantes are with Departamento de Electricidad y Electrónica, University of the Basque Country UPV/EHU, 48940 Leioa, Spain (e-mail: aitziber.anakabe@ehu.eus, juanmari.collantes@ehu.eus).

V. Armengaud is with the Centre National d'Etudes Spatiales (CNES), Toulouse Space Centre, 31401 Toulouse, France (e-mail: vincent.armengaud@cnes.fr).

the complex conjugate poles resulting from the Floquet repetitions will locate extremely close to the right half plane. The low observability of these poles together with their small real part (in the order of kHz to MHz) generates a certain level of uncertainty in the results of pole-zero identification around $f_{in}$, jeopardizing the assessment of PA stability.

In principle, a solution to overcome these difficulties is to identify directly the real poles instead of the Floquet repetitions. However, commercial tools for pole-zero stability analysis have several features that are specifically designed for the detection of resonances associated to complex conjugate poles, like overfitting detection and automatic order selection algorithms [16], [17]. In this paper, we will analyze in detail the impact of the model's long-term memory parameters on the pole-zero identification results. To reduce the ambiguity in the stability assessment, an improved algorithm for stability analysis of non-resonant frequency responses will be proposed.

The paper is organized as follows. Section II is dedicated to analyze the effect of the long-term memory parameters on the pole-zero identification around the fundamental frequency $f_{in}$. The study is illustrated through the analysis of a MMIC Doherty amplifier simulated in Advanced Design System (ADS) [26]. In Section III, an automatic identification process with an overfitting detection specifically adapted for non-resonant frequency responses is proposed and tested on the same MMIC Doherty amplifier design. The proposed approach is applied to a medium PA built in micro-strip technology with two paralleled Cree GaN HEMT devices in Section IV. Monte-Carlo (MC) stability analyses are used to show the suitability of the solutions proposed to improve the identification of transistor's slow dynamics and to avoid confusing it with potential instabilities. Finally, simulation times are discussed in Section V where the basic steps of the methodology are recapitulated.

## II. POLE-ZERO IDENTIFICATION AROUND THE FUNDAMENTAL FREQUENCY

The analysis of D-type bifurcations with pole-zero identification is usually carried out by detecting a pair of critical complex conjugate poles at the fundamental frequency $f_{in}$ [23], [24]. The stability analysis can be performed at any node or branch of the circuit, generally as close as possible to the active devices. In theory, all nodes and branches of the same circuit share the same poles, but differences might arise due to lack of observability. Some dynamics of the circuit might not be detectable from certain nodes or branches of the circuit or they might be detected with low sensitivity [27]. This reduced observability results in the identification of poles quasi-compensated by zeros of very similar values. The identification of critical poles with low observability at the fundamental frequency requires a thorough stability analysis to determine their origin.

As an example, a stable Doherty MMIC amplifier at the fundamental frequency of $f_{in}$ = 10.95 GHz with output power at saturation of 37.5 dBm is analyzed in simulation (Fig. 1). The design in Fig. 1 makes use of two UMS GH25 (0.25 μm gate length) AlGaN/GaN HEMT transistors. Access inside the UMS library models is restricted, which is very common in commercial transistor models but it includes electro-thermal and trapping parameters [2], [3].

The result of a pole-zero identification at $P_{in}$ = 20 dBm provides the pole-zero map of Fig. 2. The frequency response is simulated using a small-signal current source connected in parallel at the output node of the carrier transistor of Fig. 1. The pole-zero analysis is carried out using a commercial tool with default parameters [17]. A pair of unstable complex conjugate poles quasi-cancelled by zeros are detected at the fundamental frequency $f_{in}$ = 10.95 GHz (Fig. 2). Nevertheless, these critical poles being quasi-cancelled, their detection is highly dependent on the frequency resolution of the response, among other parameters.

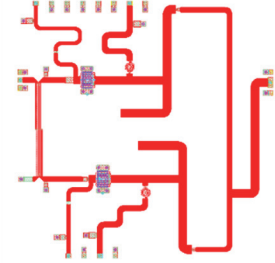

**Fig. 1.** Layout of the Doherty MMIC amplifier.

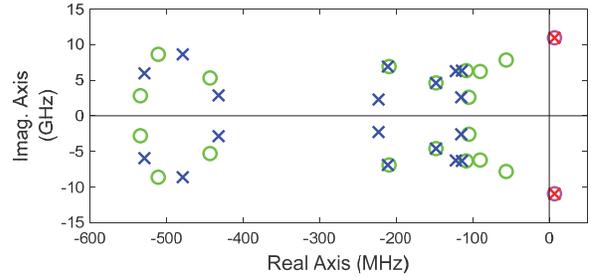

**Fig. 2.** Pole-zero map of a large-signal frequency response obtained at the output node of the carrier transistor for $P_{in}$ = 20 dBm and $f_{in}$ = 10.95 GHz.

A MIMO (Multiple-Input Multiple-Output) analysis can be performed at various significant nodes and branches of the design to detect a node or branch with better observability of the poles around $f_{in}$. To quantify the observability from the different ports, a residue analysis based on the normalized factor $\rho_k$ can be carried out [27]. The normalized factor $\rho_k$ (1) quantifies the relative effect of a pair of complex conjugate poles $p_k, p_k^*$ on the identified transfer function $H(s)$.

$$\rho_k = \frac{\left|H^k(j\omega_r)\right|}{\left|H(j\omega_r) - H^k(j\omega_r)\right|} \quad (1)$$

where $\omega_r = \sqrt{b^2 - a^2}$ is the resonance frequency of the complex conjugate pole pair $p_k = a \pm bi$ and $H^k(s)$ is a second order transfer function that represents the contribution of poles $p_k, p_k^*$ to $H(s)$:

$$H^k(s) = \frac{r_k}{s - p_k} + \frac{r_k^*}{s - p_k^*} \quad (2)$$

where the $r_k$, $r_k^*$ parameters in (2) are the residues of the poles $p_k$, $p_k^*$. High values of $r_k$, $r_k^*$ indicate that the $p_k$, $p_k^*$ poles have a significant effect on the transfer function $H(s)$, which correlates to a high observability from the observation port. Since the magnitude of $H(s)$ can have a very wide range of values, the normalization in (1) is necessary to quantify this effect. Thus, (1) represents the magnitude of the $H^k(s)$ normalized with respect to the magnitude of the transfer function $H(s)$ from which the contribution of the poles $p_k$, $p_k^*$ has been subtracted. Both magnitudes are evaluated at the resonance frequency $\omega_r$. It is conventionally accepted that $\rho_k$ values higher than 1 imply poles with reasonable high observability, while poles with $\rho_k < 0.01$ are often eliminated to reduce overfitting, as indicated in [30].

Eight frequency responses are obtained by introducing sequentially small-signal current sources in parallel at input and output nodes and small-signal voltage sources in series at the input and output branches of both transistors of the Doherty amplifier [28]. The poles at $f_{in}$ appear quasi-compensated by zeros for all the analyzed observation ports, or equivalently, as shown in Fig. 3, all the calculated normalized $\rho_k$ factors (1) are small (the maximum $\rho_k$ value in Fig. 3 is 0.18) indicating low observability at all nodes and branches [27].

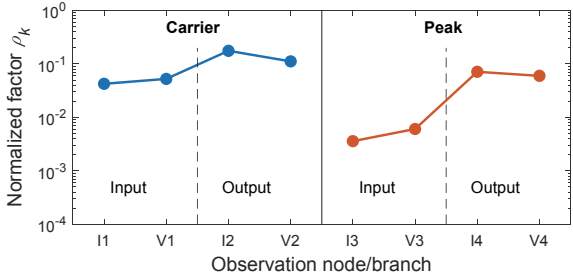

**Fig. 3.** Residue analysis of a MIMO identification of eight frequency responses obtained at the input and output nodes and branches of each transistor of the Doherty amplifier for $P_{in}$ = 20 dBm and $f_{in}$ = 10.95 GHz.

In conclusion, in this example we find critical poles at $f_{in}$, that cannot be detected clearly (with high observability) at any node or branch. As a result, the identification results presented in this section for the Doherty amplifier are inconclusive. In this situation, circuit designers may be puzzled to conclude on the stability margins of the circuit.

Actually, the presence of these poles at $f_{in}$ may be linked to the long-term memory parameters (self-heating and trapping) of the non-linear model. Nonlinear transistor models need to take into account trapping and self-heating effects for a correct prediction of transistor's long-term memory effects. An example of an electro-thermal model of an AlGaN/GaN HEMT is given in Fig. 4 [2], [8]. Self-heating is usually modeled by one or several RC cells representing the thermal impedance of the device. In the electrical simulator, the dissipated power is modeled as a current source while the output voltage across the thermal circuit represents the temperature increment (as shown in green in Fig. 4). In AlGaN/GaN HEMTs, drain-lag trapping can be modeled as a transient in the control voltage $V_{GS}$ (as shown in blue in Fig. 4) [8]. Capture and emission time constants of the traps are also modeled through RC circuits.

Capture time constants are usually very short, in the order of nanoseconds, while emission time constants are much larger. In some cases, these time constants are selected empirically to improve convergence in the harmonic-balance simulator and they can be different from the physical ones [3]. These RC networks introduce real poles that are very close to the x axis of the complex plane (long time constants) when compared with the frequencies involved in the amplifier dynamics.

For intellectual property protection, transistor manufacturers usually do not provide access or detailed information about the internal structure of their models. The electrical networks modeling these long-term memory effects are not accessible and their observability from the transistor external ports is very low. As a result, this dynamics is hardly detectable in a pole-zero analysis from the frequency responses obtained at the access nodes or branches of the transistors. If detected, we can expect the real poles associated with these long time constants to appear quasi-cancelled by near zeros.

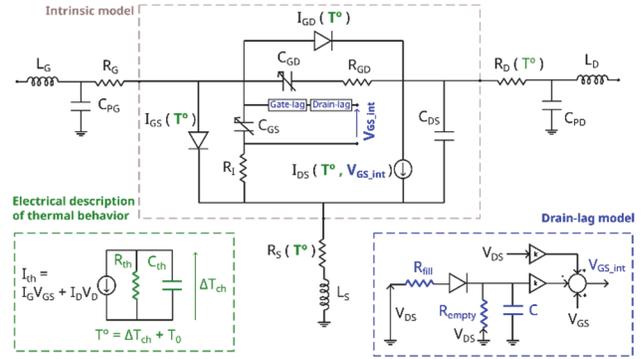

**Fig. 4.** Structure of a nonlinear model that models the thermal behavior (green) and the circuitry that models the trapping effects (blue) based on the transistor models in [2], [8].

A thorough analysis of their effect will require the identification of real poles at low frequencies, due to the long time constants involved. To that end, optimized identification procedures of real poles at low frequencies are proposed and described in the next section.

### III. OPTIMIZED DETECTION OF REAL POLES WITH LONG TIME CONSTANTS AND LOW OBSERVABILITY

In pole-zero identification for stability analysis, automatic order selection of the transfer function is essential since the order is always unknown *a priori* [13], [29]. The commercial automatic pole-zero identification algorithm in [17], [30] starts from an even number of initial complex conjugate poles that is related to the number of resonances estimated in the frequency response. Then, these poles are reallocated via different frequency domain identification algorithms (least squares or Vector Fitting [25], [31], [32] for instance). A maximum value for the phase error along the analyzed frequency band is established as fitting goal condition. If this goal is not achieved, the transfer function order is increased adding another pair of complex conjugate poles and the process is repeated until the phase error condition is met. The initial set of complex conjugate poles is recommended to be linearly distributed over the frequency response bandwidth as in [31]. Additional divide-and-conquer algorithms can also be used to speed up the identification process [29].

The use of complex conjugate poles for initializing the algorithm makes it better suited for the identification of resonant peaks in the frequency response. However, transfer functions with slow real poles have to be analyzed at very low frequencies (compared to the amplifier operation frequencies) and are non-resonant. Besides, non-resonant responses can suffer more easily from overfitting of real poles, which is the main danger of automatic order selection of the transfer function in an identification process. In the following, we propose an automatic order selection algorithm for a more efficient detection of real poles. The two main characteristics of this new in-house solution are described in the following.

First, the automatic identification algorithm will start with an initial order of $N = 1$. A stable real pole is initially located at the center frequency of the analyzed frequency response. Then, this real pole is reallocated via a frequency-domain identification algorithm. A maximum value for the phase error along the analyzed frequency band, set as default to 0.5º, is established as fitting goal condition. Each time the maximum phase error condition is not met, the transfer function order is increased with the addition of another real pole. The new set of initial poles is then a combination of the results of the previous iteration plus a new stable real pole at the center frequency value of the response. These new initial poles are again reallocated. The entire process is carried out iteratively until phase error condition is met with the final set of reallocated poles. Note that, in principle, an initial pair of real poles could also be turned into a pair of complex conjugate poles if the identification algorithm needs it for the fitting.

Second, the normalized factor $\rho_k$ in (1) is adapted for the overfitting detection of real poles. The original goal of the normalized $\rho_k$ factor was to quantify the importance of a pair of complex conjugate poles in the identified transfer function [27]. That information can be used later to eliminate extra poles due to overfitting, but also to analyze which are the nodes or branches with better observability and controllability [13]. In order to adapt (1) to the effect of a real pole, the expression in (2) has to be substituted by the first-order transfer function in (3) evaluated at $\omega_r = a$, with $a$ being the cut-off frequency of the real pole $p_k = a$:

$$H^k(s) = \frac{r_k}{s - p_k} \quad (3)$$

A third factor can further improve the detection of the real pole associated with the long-term memory parameters of the transistor model. This is not related to the identification process but to the simulation of the frequency response. Due to the topology of the electro-thermal and trapping models, the drain access of the transistor will have better observability of the slow real poles than the gate access. Generally, the model element that presents a more relevant dependence with temperature (as well as with drain-lag trapping, obviously) is the drain current source. In addition, a better observability is obtained for the series connection of the small-signal voltage source than the parallel connection of the current probe. This can be easily explained since current probes are connected in parallel at the access nodes of the transistor. At the low frequencies of the analysis, there is a low impedance path from the connecting node to ground through the bias networks. This low impedance path also reduces the observability from that node. This is shown in Fig. 5, where the $\rho_k$ factor of the real poles is computed for a MIMO analysis [27] of four different frequency responses obtained with small-signal current or voltage sources introduced at the input and output nodes and branches (respectively) of the carrier transistor in the Doherty design. We can observe that, although lower than 1 in the four cases (indicating low observability in general), the $\rho_k$ factors at the output of the transistors are larger than at the input. In conclusion, the best possible observability of the real poles associated with these long-term memory effects will be achieved with a small-signal voltage source connected in series at the output branch of the transistor. In the following, all the identifications presented in this manuscript of slow real poles are obtained with this configuration.

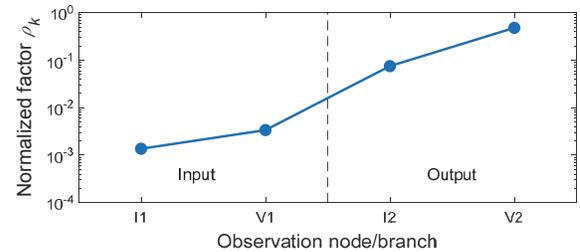

**Fig. 5.** Residue analysis of a MIMO identification of four non-resonant frequency responses obtained at the input and output nodes and branches of the carrier transistor for $P_{in} = 20$ dBm. The highest $\rho_k$ value indicates that higher observability of the real poles is obtained at the drain branch.

MC sensitivity analyses are commonly carried out to determine the robustness of the performances of a design [5], [33], including its stability margins [34]. To demonstrate the benefits of detecting the real poles instead of the complex-conjugate poles at the fundamental frequency in these cases, a large-band MC simulation with 251 iterations of the Doherty amplifier for $P_{in} = -6$ dBm has been carried out by introducing a small-signal voltage source in series at the drain branch of the carrier transistor. A dispersion of 5 % (Gaussian distribution) has been applied to all circuit components. This represents a typical MC analysis carried out by designers to analyze the complete stability margins of the PA with a single simulation in a large bandwidth.

On the one hand, the high-frequency section of the simulated response (10 MHz to 11.2 GHz) is identified with the commercial automatic pole-zero identification algorithm and overfitting detection of complex-conjugate poles in [17], [30]. The identified poles are plotted in red (unstable) and green (stable) in Fig. 6(a). On the other hand, the low-frequency section of the simulated response (50 kHz to 10 MHz) is identified with the proposed automatic algorithm for non-resonant frequency responses and overfitting detection of real poles. The identified real poles are plotted in blue in Fig. 6(a). In both identifications (for low and high-frequency sections) poles with a $\rho_k < 0.01$ have been eliminated to reduce overfitting, and zeros are not plotted to increase the visibility of the dispersion of the poles. Note that this limit is generally a good compromise to remove overfitting poles in stability analyses [30].

We can observe in Fig. 6(a) that multiple unstable poles are detected at $f_{in}$, due to the lack of observability at the fundamental frequency discussed in Section II. On the contrary, all the identified real poles in Fig. 6(a) are stable and have low scattering, as can be verified when plotting a zoom on the real poles, Fig. 6(b). The difference between the scattering of the poles at $f_{in}$ and the real poles is considerable. It is important to note that both, real poles and poles at $f_{in}$ correspond to the same Floquet multiplier. Results from Fig. 6 show that analyzing the stability margins of these critical poles at the fundamental frequency with such a wide frequency band is not practical nor time efficient. In practice, analyzing a low-frequency response is sufficient to conclude that, for this design, no D-type bifurcation is expected.

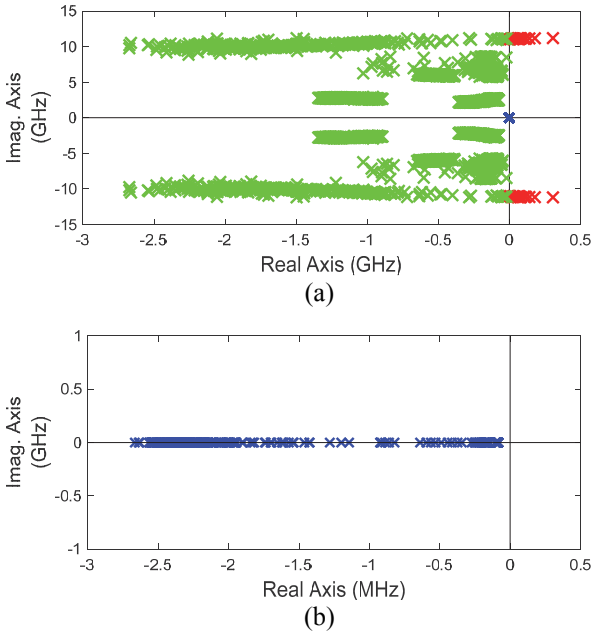

**Fig. 6.** (a) Two superimposed MC pole maps of 251 iterations are plotted for $P_{in}$ = - 6 dBm. The green and red poles represent the pole map of the wide-band frequency responses (10 MHz to 11.2 GHz). The blue poles represent the pole map for the non-resonant frequency responses (50 kHz to 10 MHz). (b) A zoom on the pole map for the non-resonant frequency responses.

## IV. APPLICATION EXAMPLE

A prototype fabricated in hybrid micro-strip technology will be analyzed to illustrate the proposed approach. The amplifier includes two GaN HEMT transistors (CHJ40010) connected in parallel. A photograph of the design is shown in Fig. 7. Access to the transistor model is also restricted for the CHJ40010 transistor, but it includes an electro-thermal model.

As in the Doherty amplifier analysis described in Section II and III, a pair of critical complex conjugate poles are detected at the fundamental frequency ($f_{in}$ = 1 GHz) when analyzing the stability of the design for increasing values of the input power, as shown in Fig. 8. The critical poles are detected as unstable for some $P_{in}$ values, but no clear input power dependency can be derived from the pole evolution versus $P_{in}$. Given that both branches of the amplifier are identical, the results obtained at the input and output of each transistor will be the same, so only results of the top branch are analyzed and plotted.

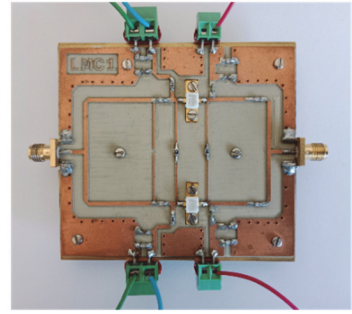

**Fig. 7.** Photograph of the hybrid microstrip GaN HEMT amplifier.

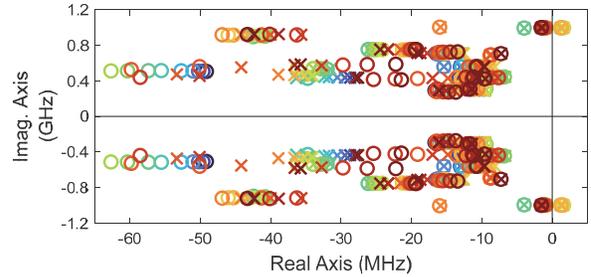

**Fig. 8.** Pole-zero map of a wide-band frequency response (250 MHz to 1.2 GHz) obtained at the output node of the top transistor for different $P_{in}$ values from 10 dBm (blue) to 26 dBm (red) with a 2 dBm step.

A MC stability analysis with 501 iterations is carried out for this amplifier in ADS ([26]). A dispersion of 5 % (Gaussian distribution) has been applied to all circuit components and transmission line widths and lengths in all the MC analyses in this section. The 501 frequency responses are simulated by introducing a small-signal voltage source at the output branch of the top transistor for $P_{in}$ = 18 dBm from 101 kHz to 1.2 GHz. A logarithm scale of 101 points per decade is used in the simulations. The poles plotted in green and red in Fig. 9(a) are the results of a pole-zero identification with a commercial tool [17] for the large-frequency portion of the simulated frequency responses: 10 MHz to 1.2 GHz. A wide dispersion of the poles at $f_{in}$ is obtained, with some of them being unstable. The identified set of slow real poles plotted in blue in Fig. 9(a) correspond to the non-resonant low-frequency portion of the simulated frequency responses: 101 kHz to 10 MHz. The real poles are identified with the automatic identification algorithm for non-resonant poles proposed in Section III. In both cases, overfitting has been eliminated by analyzing the $\rho_k$ values, and zeros are not plotted to increase the visibility of the variability of the poles. Although they appear almost completely superimposed, 501 stable real poles are plotted in Fig. 9(a). They are much less scattered than the set of complex conjugate poles identified at $f_{in}$. As determined in Section II, the real poles in Fig. 9(a) are related to the electro-thermal model of the transistor. The real poles and the complex-conjugate poles at the fundamental frequency correspond to the same Floquet multiplier. This can be confirmed by Fig. 9(b) where the same simulations and identifications are repeated with the electro-thermal model of the transistor disabled. The results of Fig. 9(b) show that, in this case, neither the real poles nor the poles at the fundamental frequency are detected.

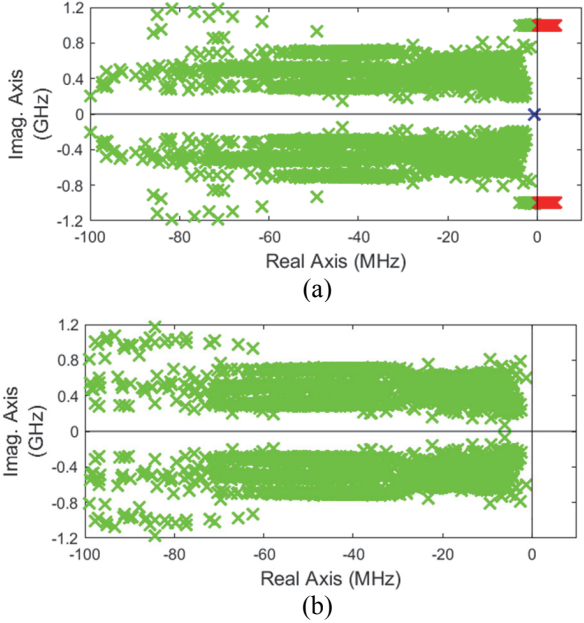

**Fig. 9.** (a) Two superimposed MC pole maps of 501 iterations are plotted for $P_{in}$ = 18 dBm with the thermal model activated. The green and red poles represent the pole map of the large-band frequency responses (10 MHz to 1.2 GHz). The blue poles represent the pole map of the low-frequency responses (101 kHz to 10 MHz). (b) The same MC analysis is repeated with the thermal model deactivated.

As shown in Fig. 9(a), MC analyses with wide-band frequency responses provide imprecise stability results at $f_{in}$. This can be also observed if the analysis is performed on narrower bands. Figs. 10 and 11 compare two MC stability analyses performed with the electro-thermal model active at low frequency (101 kHz to 5 MHz) and around $f_{in}$ (0.9 GHz to 1.1 GHz) respectively. 51 frequency points have been used in both cases. The MC analyses are performed for various input power values. The magnitude of the frequency response of one MC iteration for each $P_{in}$ value is plotted in Figs. 10(a) and 11(a). The dispersion of the real poles for the 501 MC iterations even for increasing values of the input power is almost nonexistent (Fig. 10(b)). On the contrary, the same analysis around $f_{in}$ provides a very different result. The critical poles are not detected for low power values and unstable poles are obtained for some iterations at high power, Fig. 11(b).

We can try to improve these results increasing the frequency resolution at the cost of higher simulation times. MC analyses have been repeated on the same bandwidths using 201 frequency points instead of 51. The results are plotted in Fig. 12. The simulation time increases considerably but the conclusions are very similar. As expected, the variability of the real poles in Fig. 12(a) is identical to the scattering shown in Fig. 10(b), almost nonexistent. However, although all the identified poles at $f_{in}$ are now stable, they still present a significant dispersion, Fig. 12(b).

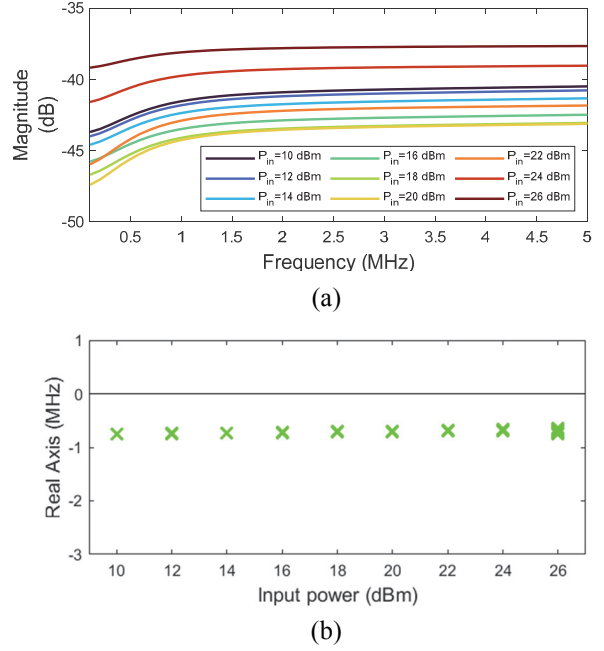

**Fig. 10.** Results of multiple low-frequency narrow-band (101 kHz to 5 MHz with 51 points) MC analyses (501 iterations for each $P_{in}$). (a) Magnitude of the frequency response of one MC iteration for each $P_{in}$. (b) Real pole versus $P_{in}$ identified for all MC iterations.

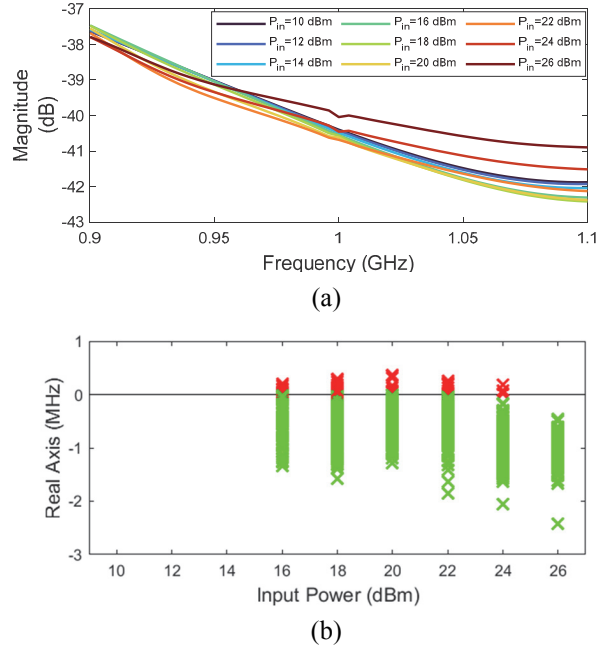

**Fig. 11.** Results of multiple narrow-band frequency responses centered around $f_{in}$ (0.9 GHz to 1.1 GHz with 51 points) MC analyses (501 iterations for each $P_{in}$). (a) Magnitude of the frequency response of one MC iteration for each $P_{in}$. (b) Real part of the critical poles at $f_{in}$ versus $P_{in}$ identified for all MC iterations.

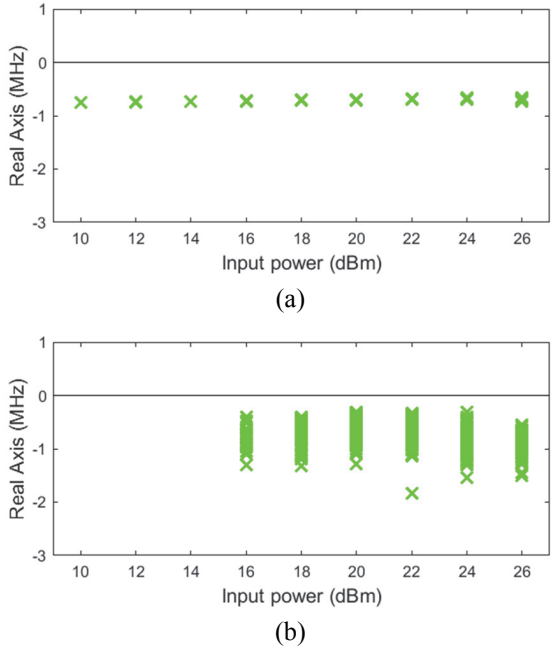

**Fig. 12.** (a) Real pole versus $P_{in}$ identified for multiple low-frequency narrow-band (101 kHz to 5 MHz with 201 points) MC analyses (501 iterations for each $P_{in}$). (b) Real part of the critical poles at $f_{in}$ versus $P_{in}$ identified for multiple narrow-band frequency responses centered around $f_{in}$ (0.9 GHz to 1.1 GHz with 201 points) MC analyses (501 iterations for each $P_{in}$).

Indeed, no D-type bifurcation is experimentally detected for the GaN HEMT amplifier in Fig. 7. As an example, the measured $P_{in}$-$P_{out}$ curve is plotted in Fig. 13 for $V_{GS}$ = -3 V and $V_{DS}$ = 20.8 V. No D-type bifurcation was detected in the laboratory for all the tested bias points and input powers additional to the results in Fig. 13. In addition, no power jump or hysteresis has been detected when raising and lowering the input power.

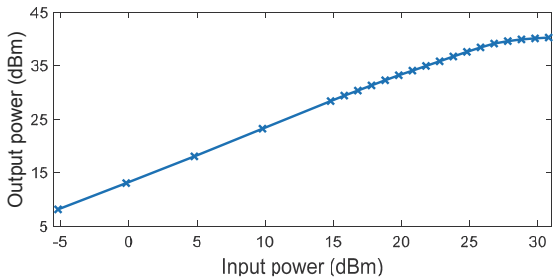

**Fig. 13.** Measured $P_{in}$-$P_{out}$ curve showing stable operation for $V_{GS}$ = - 3 V and $V_{DS}$ = 20.8 V.

## V. DISCUSSION

Results in the previous section have shown that the long-term memory parameters of the model introduce an uncertainty in the identification around the fundamental frequency that can confuse large-signal stability conclusions.

Exhaustive stability assessment of power amplifiers require large-signal stability analysis versus variations of circuit's parameters (either parametric or MC simulations). These are long analyses in which the simulation of the frequency responses in harmonic balance (large signal-small signal) is by far more time-consuming than the identification process. As an example, the simulation in ADS of the 501 frequency responses for the analysis of Fig. 9(a) requires 41 minutes (on a computer with a 12[th] Gen Intel(R) Core(TM) i7 processor with 2.69 GHz and 32 GB of RAM). The identification process of the 501 high frequency responses with [17] requires less than 30 s, while the identification of the 501 low-frequency responses with the proposed in-house algorithm takes 13 s. Thus, the simulation time in the harmonic-balance simulator is the main factor in the total time required for the analysis.

Figs. 10 to 12 showed that we could always improve pole detection around the fundamental frequency $f_{in}$ by simulating with finer frequency resolution. However, this has the cost of excessively increasing the simulation time required in ADS to obtain the frequency responses. The simulation time for the results in Fig. 12(b) is 3 hours. However, better results with low dispersion are obtained for the real poles in Fig. 10(b) with a simulation time of less than one hour.

In view of these results, the following methodology can be proposed for the large-signal stability analysis based on pole-zero identification.

First, the stability analysis of the amplifier is performed conventionally. This means that the required electrical simulations for parametric stability analyses or for MC stability analyses are carried using a frequency resolution consistent with the wide frequency bands that need to be covered, including the fundamental frequency. This allows dealing with affordable simulation times.

Second, if poles at the fundamental frequency are found, then, these stability analyses need to be complemented with an analysis of the non-resonant low-frequency band using a small-signal voltage source connected at the output branch of the transistor. The simulation of the low-frequency band can be performed with a reasonable frequency resolution according to its bandwidth, which again allows moderate execution times. The proposed algorithm for automatic identification of non-resonant responses can be used to identify the low-frequency responses, together with the overfitting detection procedure. This complementary analysis will help to conclude on the nature and stability of the real poles, reducing the uncertainties found at the fundamental frequency.

## VI. CONCLUSION

The effect of long-term memory parameters on the stability analysis of power amplifiers has been analyzed. In large-signal analysis, this slow dynamics yields a complex-conjugate pair of poles with very small real part at the fundamental frequency due to the periodic Floquet repetitions. These poles have low observability, which complicates the study of amplifier stability margins and increases the chances of confusion with an instability associated to a D-type bifurcation. In these cases, the stability margins are better analyzed by identifying the slow real poles at low frequencies. Specific procedures have been proposed for the automatic order selection of the transfer function and for the overfitting reduction of identified non-resonant frequency responses associated with the slow real poles. The benefits of the proposed analyses are illustrated in two amplifiers designed with commercial transistor models that include long-term memory parameters but have no access to the model topology.

ACKNOWLEDGMENTS

The authors of the UPV/EHU would like to thank the French Space Agency (CNES) for the design of the Doherty amplifier, developed in a common project between the two institutions.

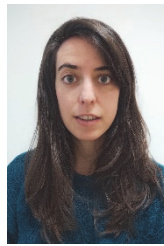

**Libe Mori** received the Ph.D. degree in electronics from the University of the Basque Country (UPV/EHU), relating to linear and nonlinear stability analysis of microwave power amplifiers in 2019. She began her carrier as a lecturer and researcher in Mondragon Unibertsitatea in 2019. In 2021 she joined the Department of Computer

Architecture and Technology, UPV/EHU. Her main research interests include stability analysis and design of microwave circuits.

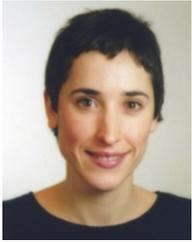

**Aitziber Anakabe** received the Ph.D. degree in electronics from the University of the Basque Country (UPV/EHU), Bilbao, Spain, in 2004. In 1999, she joined the Electricity and Electronics Department, UPV/EHU, where she was involved with the stability analysis of nonlinear microwave circuits. In 2004, she joined the French Space Agency (CNES), Toulouse, France, as a Post-Doctoral Researcher. In 2005, she rejoined the Electricity and Electronics Department, UPV/EHU, where, since 2005, she has been an Associate Professor. Her research deals with nonlinear analysis and modeling of microwave circuits and measurement techniques.

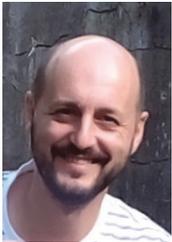

**Juan-Mari Collantes** received the Ph.D. degree in electronics from the University of Limoges, France, in 1996. Since February 1996, he has been an Associate Professor with the Electricity and Electronics Department, University of the Basque Country (UPV/EHU), Bilbao, Spain. In 1996 and 1998 he was an Invited Researcher with Agilent Technologies (formerly the Hewlett-Packard Company), Santa Rosa, CA. In 2003, he was with the French Space Agency (CNES), Toulouse, France, where he was involved with power amplifier analysis, simulation, and modeling. His areas of interest include nonlinear analysis and design of microwave circuits, microwave measurement techniques and noise characterization.

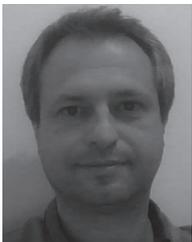

**Vincent Armengaud** received his Ph. D. in Electronics from the University of Limoges (France) in 2008. He has first worked for Thales Alenia Space France on MMIC design. In 2012 he joined the French Space Agency (CNES), Toulouse, France. His main research interests are the low noise amplifiers and medium level circuits.